\documentclass[useAMS,usenatbib,usegraphicx]{mn2e}
\def\sr{\hbox{IGR\,J18293--1213}}
\def\integral{\textit{INTEGRAL}}
\def\nustar{\textit{NuSTAR}}
\def\chandra{\textit{Chandra}}
\def\swift{\textit{Swift}}
\def\Chi2{$\chi^2$}
\def\msol{\,M$_\odot$}
\def\rsol{\,R$_\odot$}
\def\ctss{\,cts\,s$^{-1}$}
\def\ergs{\,erg\,s$^{-1}$}
\def\phcms{\,ph\,cm$^{-2}$\,s$^{-1}$}

\def\sig{\,$\sigma$}

\def\nh{$N_{\rm H}$}

\usepackage[x11names]{xcolor} 
\title[IGR\,J18293--1213 is an eclipsing CV]{IGR\,J18293--1213 is an eclipsing Cataclysmic Variable}
\author[M. Clavel et al.]{M.\ Clavel,$^{1}$\thanks{E-mail:
maica.clavel@ssl.berkeley.edu (MC)} J.~A.\ Tomsick,$^{1}$ A.\ Bodaghee,$^{2}$ J.-L. Chiu,$^{1}$ F.~M.\ Fornasini,$^{1,3}$ J.\ Hong,$^{4}$ \newauthor
R.\ Krivonos,$^{5}$ G.\ Ponti,$^{6}$ F.\ Rahoui,$^{7,8}$ and D.\ Stern$^{9}$\\
$^{1}$ Space Sciences Laboratory, 7 Gauss Way, University of California, Berkeley, CA 94720-7450, USA\\
$^{2}$ Georgia College, 231 W. Hancock St., Milledgeville, GA 31061, USA\\
$^{3}$ Astronomy Department, University of California, 601 Campbell Hall, Berkeley, CA 94720, USA\\
$^{4}$ Harvard-Smithsonian Center for Astrophysics, 60 Garden St., Cambridge, MA 02138, USA\\
$^{5}$ Space Research Institute, Russian Academy of Sciences, Profsoyuznaya 84/32, 117997 Moscow, Russia\\
$^{6}$ Max-Planck-Institute f\"ur Extraterrestriche Physik, Gissenbachstrasse, 85748 Garching, Germany\\
$^{7}$ European Southern Observatory, Karl Schwarzchild-Strasse 2, 85748 Garching bei M\"unchen, Germany\\
$^{8}$ Department of Astronomy, Harvard University, 60 Garden Street, Cambridge, MA 02138, USA\\
$^{9}$ Jet Propulsion Laboratory, California Institute of Technology, Pasadena, CA 91109, USA
}

\begin{document}

\date{Accepted 2016 May 27. Received 2016 May 24; in original form 2016 April 12}

\pagerange{\pageref{firstpage}--\pageref{lastpage}} \pubyear{2016}

\maketitle

\label{firstpage}

\begin{abstract}
Studying the population of faint hard X-ray sources along the plane of the Galaxy is challenging because of high-extinction and crowding, which make the identification of individual sources more difficult. \sr\ is part of the population of persistent sources which have been discovered by the \textit{INTEGRAL} satellite. We report on \nustar\ and \swift/XRT observations of this source, performed on 2015 September~11.  We detected three eclipsing intervals in the \nustar\ light curve, allowing us to constrain the duration of these eclipses, $\Delta t= 30.8^{+6.3}_{-0.0}$\,min, and the orbital period of the system, $T=6.92\pm0.01$\,hr. Even though we only report an upper limit on the amplitude of a putative spin modulation, the orbital period and the hard thermal Bremsstrahlung spectrum of \sr\ provide strong evidence that this source is a magnetic Cataclysmic Variable (CV). Our \nustar\ and \swift/XRT joint spectral analysis places strong constraints on the white dwarf mass $M_{\rm wd} = 0.78^{+0.10}_{-0.09}$\msol. Assuming that the mass to radius ratio of the companion star $M_\star/R_\star = 1$ (solar units) and  using $T$, $\Delta t$ and $M_{\rm wd}$, we derived the mass of the companion star $M_\star = 0.82\pm0.01$\msol, the orbital separation of the binary system $a=2.14\pm0.04$\rsol, and its orbital inclination compared to the line of sight  $i=(72.2^{+2.4}_{-0.0})\pm1.0$\degr.
\end{abstract}

\begin{keywords}
stars: individual (\sr), white dwarfs, X-rays: stars
\end{keywords}

%%%%%%%%%% Section 1: Introduction
\section{Introduction}

The \integral\ mission \citep{winkler2003} has been surveying the hard X-ray sky for more than a decade, detecting close to a thousand sources \citep[see][for the most recent catalog]{bird2016}. About half of these detections are in the direction of the Galactic plane \citep[$\vert b\vert <17.5$\degr,][]{krivonos2012} and among those, two dozen sources are reported as unidentified persistent sources, with a 17--60\,keV flux below the 0.7\,mCrab limit of the \integral\ survey \citep[e.g.][]{lutovinov2013}. The goal of the `Unidentified \integral\ sources' legacy program conducted by the \textit{Nuclear Spectroscopic Telescope Array} (\nustar) is to take full advantage of the better sensitivity and the higher spatial resolution of this instrument to investigate these faint persistent sources using detailed spectral and variability analyses. These individual identifications will help to characterize the population of faint hard X-ray sources in the Galaxy by improving the completeness of the current sample.

Faint Galactic sources with hard X-ray spectra are likely to be either Cataclysmic Variables (CVs), Low Mass X-ray Binaries (LMXBs) or High Mass X-ray Binaries (HMXBs). These three types of sources are accreting compact binaries. However, the nature of the accretion onto the compact object, a white dwarf for CVs and a neutron star or a black hole for LMXBs and HMXBs, is responsible for different orbital parameters and different emission processes. These differences can be used to discriminate between these three categories of hard X-ray sources \citep[see][for detailed reviews on their properties]{kuulkers2006,tauris2006}.  

IGR\,J18293--1213 has been reported in successive \textit{INTEGRAL}/IBIS catalogs \citep{krivonos2010,krivonos2012,bird2016}.  A \swift/XRT position and spectrum were also obtained for this source \citep{landi2010} and its \chandra\ and near-infrared counterparts have been identified \citep{karasev2012}. However, these observations did not allow for a conclusive identification of the nature of this source. \sr\ was therefore part of the list proposed for the `Unidentified \integral\ sources' \textit{NuSTAR} legacy survey and was the first one to be observed with \nustar\ and \swift/XRT as part of this program. In the present paper, we describe the corresponding observations and data reductions (Sec.\,\ref{sec:dataAnalysis}). The new constraints we obtained using both the light curve (Sec.\,\ref{sec:eclipse}) and the spectra (Sec.\,\ref{sec:spec}) of \sr\ are sufficient to identify this source as an Intermediate Polar (IP, a subcategory of CVs), and to constrain the orbital parameters of this system (Sec.\,\ref{sec:sys}). These results are discussed in Sec.\,\ref{sec:conclu}.

%%%%%%%%%% Section 2: Data reduction
\section{Observations and data reduction}
\label{sec:dataAnalysis}
The \nustar\ legacy program observations of \sr\ were performed on 2015 September~11, and are summarized in Table\,\ref{tab:obs}. 
\begin{table*} 
        \centering
        \caption{\nustar\ legacy program observations of \sr.}
        \label{tab:obs}
        \begin{tabular}{l c c c c}
        \hline \hline
        Mission/Instrument & Obs. ID & Start Time (UT) & End Time (UT) & Clean exposure (ks)  \\
        \hline
        \nustar/FPMA\&B & 30161002 & 2015-09-11 10:56:08 & 2015-09-12 04:06:08 & 25.71$^{\rm *}$ \\
        \swift/XRT & 00081763001 & 2015-09-11 16:32:55 & 2015-09-11 18:20:54 & \hspace{4pt}1.89\hspace{4pt} \\
        \hline \hline
        \end{tabular}
        
		{\em $^{\rm *}$} The \nustar\ effective exposure used for the spectral analysis (excluding the eclipse intervals) is 23.07\,ks.\\
    
\end{table*}
Our data set is composed of two observations made with the two co-aligned X-ray telescopes on board the \nustar\ mission, both of which cover the 3--79\,keV energy range \citep{harrison2013}. We also obtained a short observation with the \swift\ \citep{gehrels2004} X-ray telescope \citep[XRT;][]{burrows2005} covering the 0.5--10\,keV energy range.
\subsection{\nustar}
We reduced the \nustar\ data using NuSTARDAS v.1.5.1 which is part of HEASOFT 6.17, setting \texttt{saamode=strict} and \texttt{tentacle=yes} in order to better remove the time intervals having an enhanced count rate due to the contamination created by the South Atlantic Anomaly (SAA).  

\subsubsection{Further cleaning of the SAA contamination}
\label{sec:saa}
The SAA contamination was not fully removed by the automated procedure and could be seen as multiple periods with increased count rates. In order to remove the corresponding periods from the Good Time Intervals (GTI), we extracted the light curve of the whole detector for both \nustar\ focal plane modules (FPMA \& FPMB), using a binning of 10\,s. We then removed from the GTI all bins which had an exposure fraction lower than 0.8 and all continuous bins which were at least 1.5\sig\ above the average count rate, simultaneously in both FPMA and FPMB. This last step was iterated twice in order to correct for any artificial increase of the average count rate created by the SAA contamination.

\subsubsection{Light curve and spectral extraction}
\label{sec:lc+spec}
We extracted the source light curves and spectra from a circular region having a 60$''$ radius and centered on the \chandra\ position of \sr\ \hbox{(${\rm R.A.}=18^{\rm h}29^{\rm m}20.16^{\rm s}$, ${\rm Dec.}=-12^\circ12'50.7''$, J2000)}\footnote{The \nustar\ detection is in agreement with the \chandra\ position, within the systematic uncertainties of a few arcseconds.}. 
The background light curves and spectra were extracted from a circular region having a 100$''$ radius and located at the other end of the source detector chip. 

We applied the barycenter correction to the photon arrival times by setting \texttt{barycorr=yes}, to create both the event files used to constrain the variability of the source, and the light curves. Unless stated otherwise, the source light curves are not background subtracted, but are presented along with the  background light curves scaled to the area covered by the source region. The source lightcurves are shown with the gaussian 1\,\sig\ error bars which lead to an underestimation of the corresponding uncertainties for small bins at low count rates (e.g.\ during the eclipse periods, Fig.\,\ref{fig:lc}, bottom panel). However, these error bars are only used for display purposes, while variability studies are mainly based on the event files (Sec.\,\ref{sec:eclipse}).

The spectra were extracted after removing the eclipse time intervals (Sec.\,\ref{sec:eclipse}) from the GTI to improve the signal-to-noise ratio. Then, each spectrum was grouped to reach at least a 5\sig\ significance in each energy bin, except for the highest energy ones for which we have a significance of 3.3\sig\ (31.72--79\,keV) and 3.2\sig\ (25.96--79\,keV), for FPMA and FPMB, respectively.

\subsection{\swift}
The \swift/XRT was operated in Photon Counting (PC) mode and the corresponding data were reduced using HEASOFT v6.17. 
For the source, we extracted a spectrum from a 30$''$ radius centered on \sr.  We also made a background spectrum using an annulus with inner and outer radii of 60$''$ and 300$''$, respectively.  We obtained a source count rate of $0.029 \pm 0.004$\ctss\ (0.5--10\,keV).  We used the most recent response matrix for a spectrum in PC mode (swxpc0to12s6\_20130101v014.rmf), and we used {\ttfamily xrtmkarf} with an exposure map to make the ancillary response file.  Finally, we grouped the spectrum by requiring bins in which the source is detected at the 3.4\sig\ level or higher except for the highest energy bin (6.3--10\,keV) which has a significance of 1.9\sig.

%%%%%%%%%% Section 3: Lightcurve & Spectrum
\section{New constraints on \sr}
\label{sec:constr}
Using the \textit{NuSTAR} light curve of \sr\ we are able to put strong constraints on the eclipse parameters (Sec.\,\ref{sec:eclipse}). This is the first time eclipses have been reported for this system. The \swift\ and \nustar\ joint spectral analysis give additional constraints on the nature and the parameters of this binary system (Sec.\,\ref{sec:spec}).

\subsection{Light curve including eclipses}
\label{sec:eclipse}
\begin{figure*}
	\centering
	\includegraphics[width=0.81\textwidth]{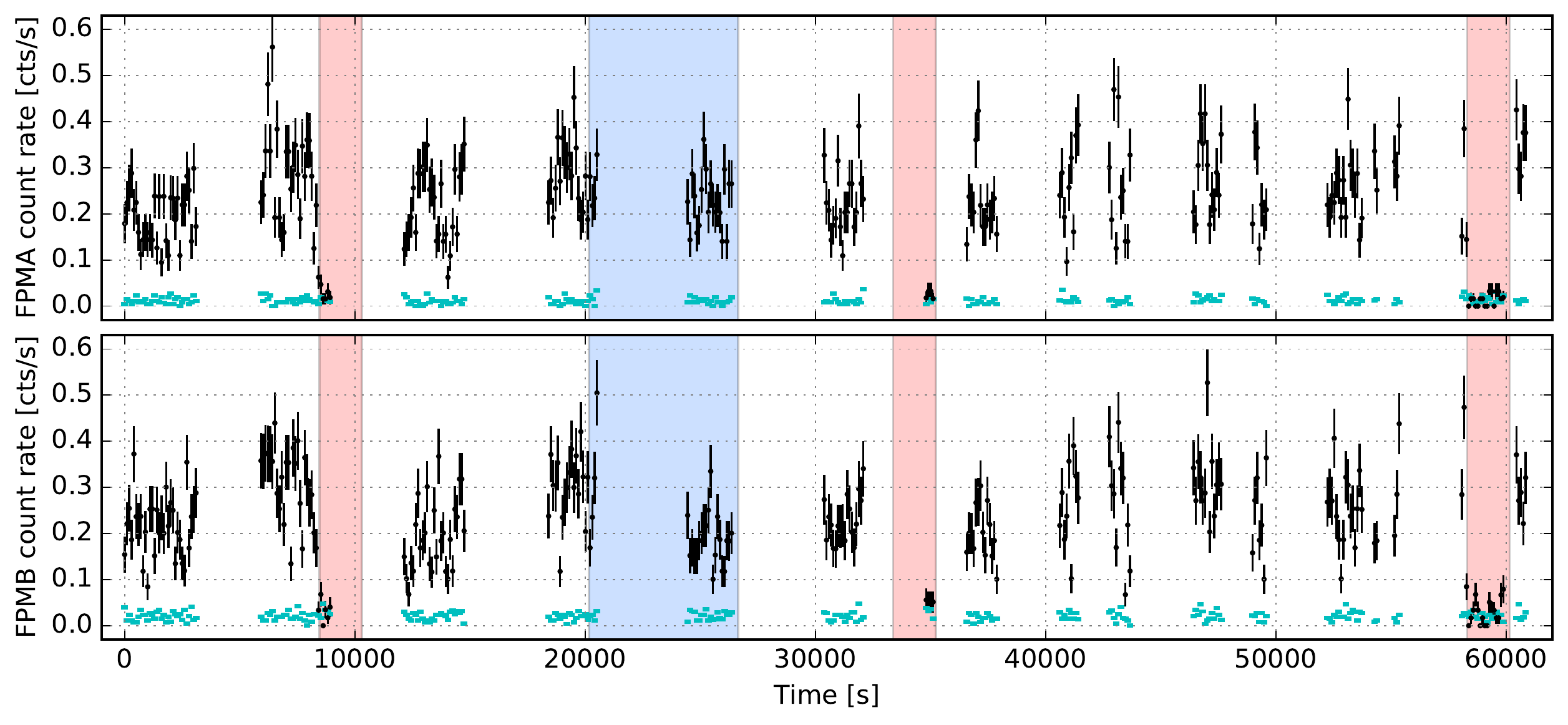}
	\includegraphics[width=0.81\textwidth]{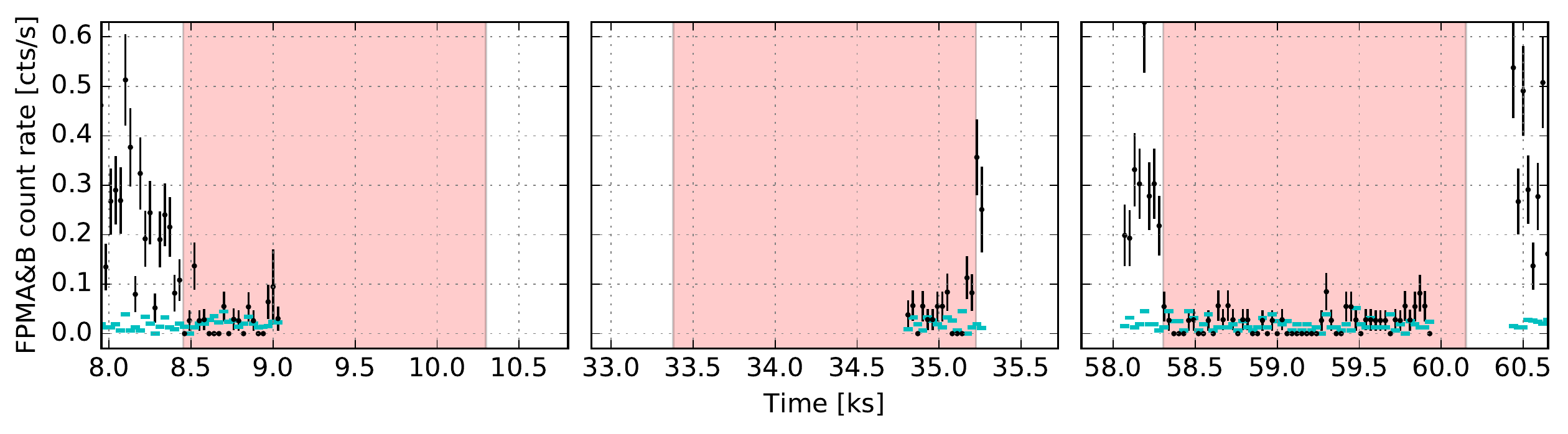}
	\caption{\nustar\ light curves of \sr\ (black) and corresponding background (cyan) as defined in Sec.\,\ref{sec:lc+spec}. Top panels: FPMA and FPMB individual light curves sampled with a bin size of 100\,s. Bottom panels: zoom-in on specific periods of the \nustar\ average light curves sampled with a bin size of 30\,s. The time reference corresponds to the first bin of our observation. The gaps in the light curves are due to the Earth occultation and/or to the SAA passages. The red shade highlights the three eclipses covered by our observations as defined by the Bayesian block analysis (Sec.\,\ref{sec:eclipse}). The blue shade highlights the simultaneous \swift/XRT observation, which does not cover any eclipse.}
	\label{fig:lc}
\end{figure*}

The source and background light curves are presented in Fig.\,\ref{fig:lc} (top panels). Within each \nustar\ module, the source count rate is compatible with a constant emission \hbox{($\sim0.25$\ctss)}, apart from three intervals during which its flux drops to the background emission level \hbox{($\sim0.016$\ctss)}. 

\subsubsection{Bayesian block analysis}
\label{sec:bb}

In order to get precise start and stop times of the eclipsing periods (i.e. independent of the light curve binning) we used the source event file including all events detected in FPMA or FPMB during their common GTI, and relied on the Bayesian block analysis described by \cite{scargle2013} and provided by P.~K.~G.\ Williams\footnote{https://github.com/pkgw/pwkit/blob/master/pwkit/bblocks.py}. This method models the continuous light curve (i.e. ignoring the numerous gaps present in our data set) with a succession of blocks having constant count rates, and it finds the optimal location for the transition times. The overall description of the light curve depends on the probability of detecting a fake extraneous block. The parameter $p_0$ is an estimation of this probability and it was set to 0.05 for our analysis. Among the ten blocks we detected, only three are compatible with the background level. The corresponding time intervals are given in Table\,\ref{tab:eclipse}, and they match the three low count-rate periods seen in the light curves (Fig.\,\ref{fig:lc}).

\begin{table} 
        \centering
        \caption{Bayesian blocks defining the eclipse intervals. The time reference is given by $t_0 + t_{bar} = 179665452{\rm\,s} + 161$\,s where $t_0$ is the starting time of the GTI given by the clock on board \nustar\ and $t_{bar}$ is the corresponding barycenter correction.}
        \label{tab:eclipse}
        \begin{tabular}{c r r}
        \hline \hline
        Eclipse block & $t_{\rm start}$ (s) &$t_{\rm stop}$ (s)\\
        \hline     
        1 & 8449\hspace{4pt} &9048$^{\rm *}$ \\
		2 & 34824$^{\rm *}$ & 35220\hspace{4pt} \\        
        3 & 58300\hspace{4pt} & 59954$^{\rm *}$  \\
        \hline \hline
        \end{tabular}
        \begin{flushleft}
		{\em $^{\rm *}$} These transitions are associated with large exposure gaps (see Fig.\,\ref{fig:lc}): they cannot be used to derive the eclipse parameters.\\
		\end{flushleft}
\end{table}

\subsubsection{Eclipse parameters}
\label{sec:tT}
The Bayesian block description ignores the observation gaps, so only the transitions detected within a continuous observation interval can be used to derive the parameters of the eclipsing signal. In particular, the three eclipses we detected are only partially covered by the \nustar\ effective exposure (Fig.\,\ref{fig:lc}, bottom panels, and Table\,\ref{tab:eclipse}). Therefore, assuming that the eclipsing signal is periodic, we used the starting times of the first and third eclipse and the ending time of the second eclipse in order to derive the eclipse properties.

The Bayesian block analysis does not provide any uncertainty for these transition times. However, in the present work these uncertainties are dominated by the systematic error linked to the shape chosen to fit the eclipse (Fig.\,\ref{fig:lc}, bottom panels). Using the unweighted light curve with 30\,s bins, we tested models including a linear transition between the detection and the non detection periods. A slow transition of about 500\,s seems to be relevant to fit the shape of the first eclipse we detected, leading to a starting time shift of about 90\,s. For the others, the transition seems to be faster (less than 100\,s) and the transition time is better constrained. However, the second eclipse may not have sufficient coverage to properly fit a linear transition, and, adding information from the third eclipse, we constrained the stopping time shift to be less than 290\,s. Therefore, we fix the systematic errors to be $t_{\rm start}$\,$^{+0}_{-90}$ and $t_{\rm stop}$\,$^{+290}_{-0}$\,s, where $t_{\rm start}$ and $t_{\rm stop}$ are the Bayesian block values listed in Table\,\ref{tab:eclipse}.

\begin{figure}
	\centering
	\includegraphics[trim = 0mm 0mm 0mm 0mm, clip, width=\columnwidth]{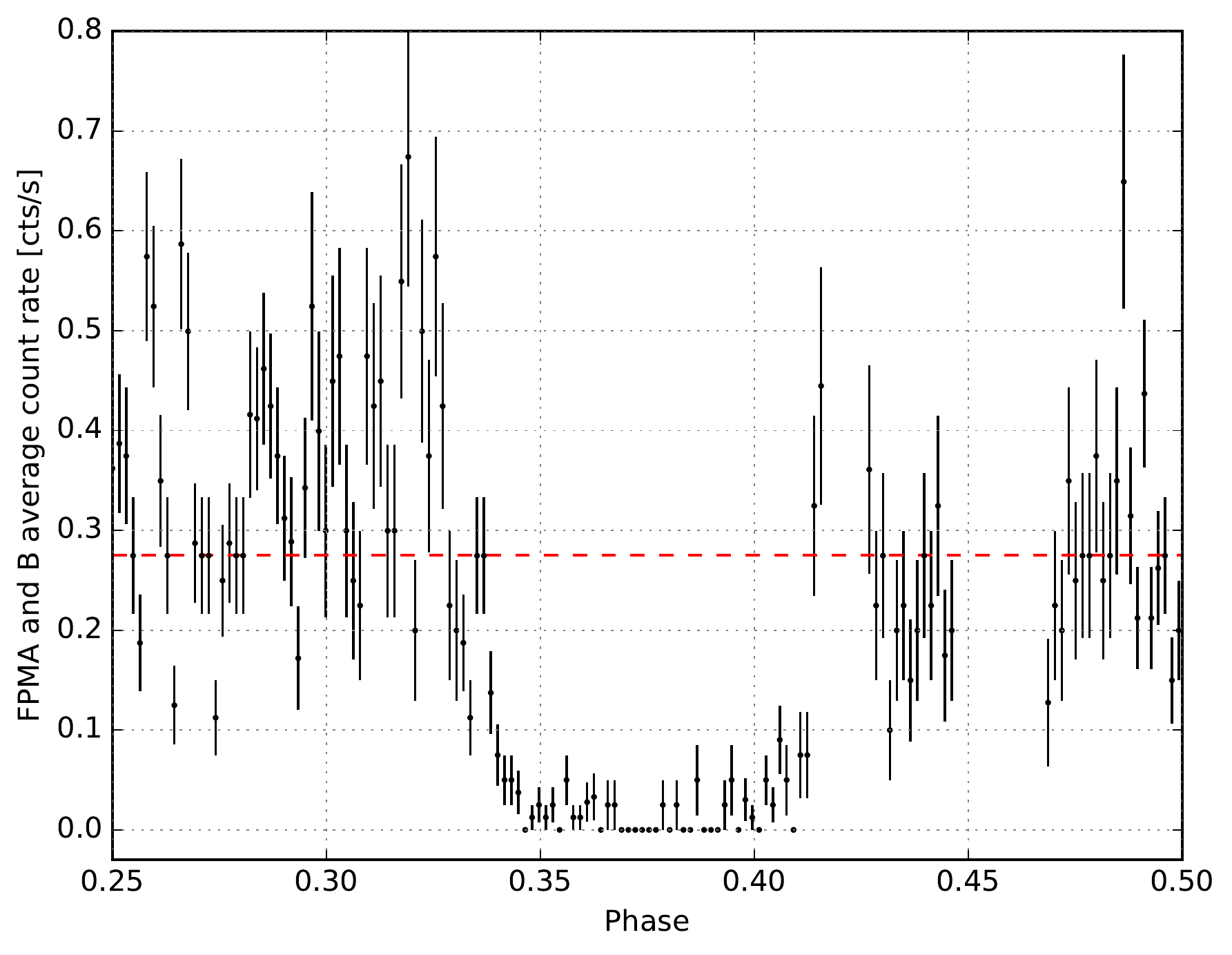}
	\caption{Profile of the eclipse obtained by folding the \nustar\ 3--79\,keV lightcurve on \sr\ orbital period, $T\sim6.92$\,hr. Each bin corresponds to a 40\,s interval and the source mean count rate is given by the dashed-red line.
	}
	\label{fig:foldlc}
\end{figure}

Using these values and their systematic errors, we obtained the duration of the eclipse, $\Delta t=30.8^{+6.3}_{-0.0}$\,min, and the signal period, $T=6.92\pm0.01$\,hr. The eclipse profile obtained by folding the light curve on \sr's orbital period $T$ is shown in Fig.\,\ref{fig:foldlc}. A shorter period is excluded by the detection of the source within at least one of the putative additional eclipse intervals, e.g.\ the detection in the 46300--47700\,s interval excludes the possibility that the period is twice as short (Fig.\,\ref{fig:lc}, top panels). 

The short orbital period we detected for \sr\ is a strong indication that this system is either a CV or an LMXB. Indeed, the majority of known CVs have orbital periods of a few hours and LMXBs of the order of a day or less, while HMXBs are generally observed with orbital periods of a few days or more \citep[e.g.][]{kuulkers2006,tauris2006}.

\subsubsection{Upper limit on spin modulation}
\label{sec:spin}
Modulations at the spin period, caused by the variation of photoelectric absorption and/or self-occultation, have been observed in several CVs. In order to search for such a short periodic signal, we made \nustar\ light curves in the \hbox{3--24\,keV} energy range using 0.05\,s time bins and combining the counts from FPMA and FPMB.  We extracted counts from the same source and background regions used for spectral analysis. We also used the same GTI as for the spectral analysis; thus, the eclipse times were removed. 

\begin{figure}
	\centering
	\includegraphics[trim = 25mm 130mm 20mm 30mm, clip, width=\columnwidth]{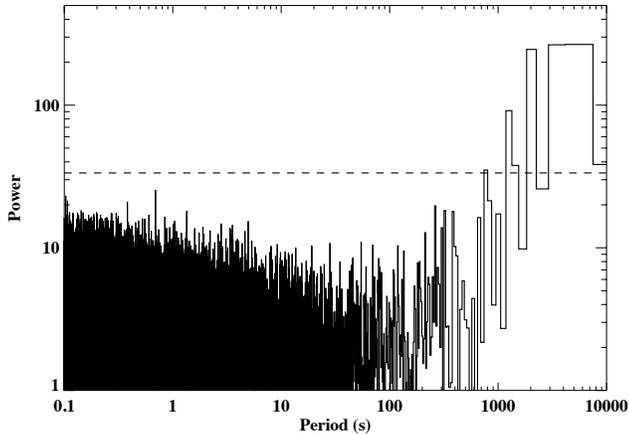}
	\caption{Periodogram obtained from the 3--24\,keV \textit{NuSTAR} lightcurve of \sr. The horizontal dashed line corresponds to a 3\sig\ threshold. 
	}
	\label{fig:zplot}
\end{figure}

We used the $Z_{1}^{2}$ (Rayleigh) test \citep{buccheri1983} to search for signals, making a periodogram extending from 0.0001\,Hz (10,000\,s) to the Nyquist frequency (10\,Hz). From 0.1 to about 750\,s, there are no signals that reach the 3\sig\ significance threshold (after accounting for trials).  Although there is a 3\sig\ peak at 769\,s, it is the first of a series of peaks that increase in significance with increasing period, and we suspect that we are simply seeing evidence for aperiodic variability (Fig.\,\ref{fig:zplot}).  

To determine the upper limit on the strength of a periodic signal, we folded the background subtracted light curve on 10,000 frequencies between 0.0001\,Hz (10,000\,s) and 1\,Hz (1\,s) using ten phase bins. For each trial frequency, we recorded the $\chi^{2}$ value obtained when fitting a constant and also the amplitude of the folded light curve in terms of the maximum count rate minus the minimum count rate divided by the sum of these two quantities.  From 1\,s to 1000\,s, 95\% of the trials have amplitudes below 9.2\%, indicating that this is the 2\sig\ upper limit on periodic signals in the 3--24\,keV band.  From 1000\,s to 10,000\,s, the highest amplitude is 19$\pm$4\% (1\sig\ error), indicating a 2\sig\ upper limit of 27\%. The implication of this upper limit will be discussed in Sec.\,\ref{sec:conclu}. 

\subsection{Spectrum compatible with a CV}
\label{sec:spec}

\begin{figure}
	\centering
	\includegraphics[width=\columnwidth]{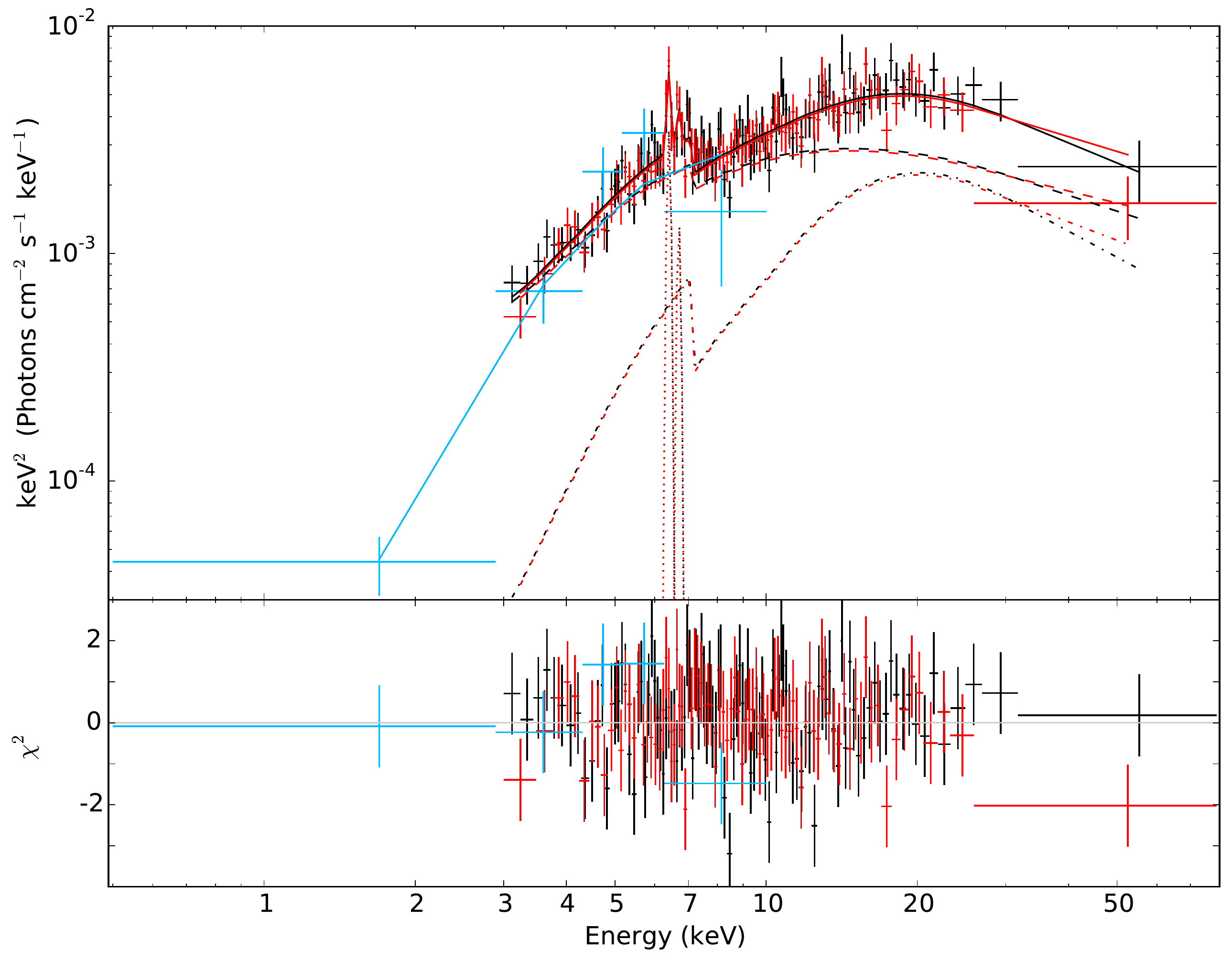}
	\caption{Unfolded spectrum (top) and residuals (bottom) obtained for \nustar\ (FPMA, black; FPMB, red) and \swift/XRT (blue) joint spectral analysis of \sr. The model we use and the parameters we obtained are given in eq.\,(\ref{eq:model}) and in Table\,\ref{tab:param}, respectively. The continuum components are the IPM model (dashes) and the corresponding reflection (dash-dots).}
	\label{fig:spec}
\end{figure}

The \nustar\ and \swift/XRT joint spectrum of \sr\ is presented in Fig.\,\ref{fig:spec}. All the model fits were performed using XSPEC v.12.9.0. A simple absorbed power-law poorly fits the data (\Chi2/d.o.f. = 277.7/193), and the residuals highlight the presence of a high-energy cutoff and of a Gaussian emission line around 6.5\,keV. Including these two additional components to the model leads to a good fit (\Chi2/d.o.f. = 185.1/190) with a photon index $\Gamma=0.4\pm0.2$. With such a hard spectrum, this source cannot be an LMXB and \sr\ is therefore likely to be a CV \citep[see Sec.\,\ref{sec:tT} and][]{tauris2006}. 

The spectrum of this category of sources is generally fitted with an absorbed Bremsstrahlung model, a Gaussian line to account for the iron-line complex around 6.5\,keV and sometimes an additional partial covering absorption component \citep[e.g.][]{suleimanov2005,mukai2015}. For \sr, a partial covering component is needed (\Chi2/d.o.f. = 298.4/191 down to 180.9/189 when \texttt{pcfabs} is added), and the Bremsstrahlung temperature given by the best fit is $kT=17.3^{+5.7}_{-3.3}$\,keV. Such a high temperature is a strong indication that this source is a magnetic CV \citep[non-magnetic CV are characterized by lower temperatures, $kT\sim1$--$5$\,keV,][]{kuulkers2006}.

Among magnetic CVs, there are two subcategories: Polars and Intermediate Polars (IPs), based on the strength of the white dwarf magnetic field, strong and intermediate, respectively \citep{kuulkers2006}. The hardness of \sr's spectrum, the relatively high luminosity of this source, its rather long orbital period and the absence of visible variability in its light curve are strong indications that \sr\ is an IP (see Sec.\,\ref{sec:conclu} for more details).

\subsubsection{IP spectral model}
\label{sec:model}

In IPs, accreted material, coming from the truncated accretion disk, follows the magnetic field of the white dwarf towards its magnetic poles. It experiences a strong shock before reaching the compact object and then cools on its way down. This column of cooling material produces the main source of X-ray radiation, and we used the spectral model \texttt{IPM} to account for its continuum component  \citep{suleimanov2005}. 
Part of this emission is reflected onto the surface of the white dwarf, and we used the XSPEC model \texttt{reflect} to represent the continuum emission created by this second process \citep{magdziarz1995}. To account for the iron fluorescence lines which are not included in the previous models, we also added two Gaussian lines to our model. Therefore, the spectral model we used is the following:

\begin{equation}
	\textsf{const}*\textsf{tbabs}*\textsf{pcfabs}*(\textsf{reflect}*\textsf{IPM}+\textsf{gauss}+\textsf{gauss}),
	\label{eq:model}
\end{equation}

\noindent
where \texttt{pcfabs} takes into account the partial absorption which can be created by the accretion flow itself, and where \texttt{tbabs} models the column density towards the source. The constant \texttt{const} accounts for imprecision in the instruments cross-calibration.

\begin{table}
        \centering
        \caption{Spectral parameters obtained by fitting model\,(\ref{eq:model}) to the \nustar\ and \swift/XRT joint spectrum of \sr. The uncertainties listed correspond to 90\% confidence intervals. The normalization constant is fixed to 1 for FPMA, and is \hbox{$0.97\pm0.04$} and $0.8\pm0.2$ for FPMB and XRT, respectively. 
        The X-ray flux of the source is $1.6$ and $0.9\times10^{-11}$\,erg\,cm$^{-2}$\,s$^{-1}$ in the 0.1--100 and 3--20\,keV range, respectively.
        }
        \label{tab:param}
        \renewcommand{\arraystretch}{1.5}
        \begin{tabular}{l c c c}
        \hline \hline
        Model & Param. & Unit & Best Fit  \\
        \hline
		tbabs & \nh & $10^{22}$\,cm$^{-2}$ &  $4.0$ (fixed) \\
		pcfabs & \nh$_{,\rm pc}$ &  $10^{22}$\,cm$^{-2}$ &  $42.8^{+19.2}_{-14.6}$ \\ 
					&  fraction & --- &  $0.70^{+0.08}_{-0.06}$\\ \hline
		reflect  & $\Omega/2\pi$ & --- & $1.0$ (fixed)\\
					& A & --- & $0.56^{+2.90}_{-0.44}$\\ 
					&A$_{\rm Fe}$& --- & $1.0$ (fixed)\\
					& cos\,$\alpha$ & --- & $>0.5$\\ \hline
		IPM & $M_{\rm wd}$ &\msol & $0.78^{+0.10}_{-0.09}$\\
					 & F$_{1-79\rm keV}$& $10^{-3}$\phcms & $2.4^{+1.3}_{-0.7}$\\\hline
		gauss & E$_{\rm Fe\,K\alpha}$ & keV & $6.4$ (fixed)\\
					   & $\sigma_{\rm Fe\,K\alpha}$ & eV & $50$ (fixed)\\
					   & N$_{\rm Fe\,K\alpha}$ & $10^{-5}$\phcms & $1.7^{+0.8}_{-0.7}$\\ \hline
		gauss & E$_{\rm Fe\,xxv}$ & keV & $6.7$ (fixed)\\
					   & $\sigma_{\rm Fe\,xxv}$ & eV & $50$ (fixed)\\ 
					   & N$_{\rm Fe\,xxv}$ & $10^{-5}$\phcms & $ < 0.13$\\ 
        \hline \hline
        \end{tabular}  
\end{table}

\subsubsection{Constraints on the parameters}
\label{sec:spec-param}
After setting the abundances to the values provided by \citet{wilms2000}, the column density towards the source was fixed to \hbox{$N_{\rm H}=4\times10^{22}$\,cm$^{-2}$}. This is the sum of the hydrogen contribution ($\sim1.3\times10^{22}$\,cm$^{-2}$, average value given by the HEASOFT tool -- Leiden/Argentine/Bonn Survey of Galactic HI) and of the molecular contribution ($\sim2.6\times10^{22}$\,cm$^{-2}$, converted from the CO map\footnote{The antenna temperature of CO at the position of \sr\ is $W_{\rm CO}=72$\,K\,km\,s$^{-1}$. We used this value and the conversion factors, $N_{\rm H_2}/W_{\rm CO}=1.8\times10^{20}$\,cm$^{-2}$\,K$^{-1}$\,km$^{-1}$\,s and $N_{\rm H}/N_{\rm H_2}$ = 2, to obtain the molecular contribution to the total column density, \nh.} provided by \citealt{dame2001}) estimated in the direction of \sr. When free to vary, this parameter is poorly constrained:  $N_{\rm H}=(4.3^{+4.8}_{-2.5})\times10^{22}$\,cm$^{-2}$, but the value obtained 
is consistent with the above estimation.

The intensity of the reflection component depends on the fraction of the X-ray flux intercepted by the white dwarf, which is represented by a solid angle scaled to unity for an isotropic source above a disk extending to infinity. When left free, this parameter reaches $\Omega/2\pi \sim 1.7$, which is not physical for the source we consider. Therefore, we fixed this parameter to~1, i.e. to the highest meaningful value. In addition, the abundance of iron, A$_{\rm Fe}$, was set to 1 compared to the abundance of elements heavier than helium, A, so the global abundance of the reflecting material can be directly compared to that of the sun.

The Gaussian line energies were fixed to 6.4 and 6.7\,keV to account for the fluorescent lines Fe\,K$\alpha$ and Fe\,\textsc{xxv}, respectively. In CVs these emission lines are observed to be relatively narrow \citep[e.g.][]{hellier2004,hayashi2011}, so we fixed their widths to 50\,eV.

The other parameters are free to vary within the default range allowed by the corresponding XSPEC models. The fit is satisfactory (\Chi2/d.o.f. = 177.6/188) and, except for the 6.7\,keV emission line, all components listed in the model are statistically required (removing the partial covering absorption or the reflection component lead to fits having \Chi2/d.o.f. = 315.8/190 and  \Chi2/d.o.f. = 213.7/190, respectively). The results are shown in Fig.\,\ref{fig:spec} and in Table\,\ref{tab:param}. The parameters of the \texttt{IPM} model are well constrained by the fit and the white dwarf mass we obtain, $M_{\rm wd} = 0.78^{+0.10}_{-0.09}$\msol, is consistent with the typical mass generally observed for white dwarfs in IPs \citep[e.g.][]{suleimanov2005,yuasa2010,hailey2016}. In our system, the free parameter $\alpha$ of the \texttt{reflect} model is the inclination of the white dwarf magnetic field compared to the line of sight. The fit only gives an upper limit $\alpha < 60$\degr\ for this parameter. Finally, of the two emission lines added to our model, only the 6.4\,keV line is significantly detected. Its equivalent width is $EW_{\rm Fe\,K\alpha}\sim160$\,eV, which is in the range of values typically reported for this type of source \citep[e.g.][]{hellier2004}. The 6.7\,keV emission line is only marginally detected in our data set, so we report an upper limit on its normalization. Removing this component from our model does not change the continuum model parameters, except for the abundance of the reflection model which is then constrained to values less than solar.

%%%%%%%%%% Section 4: Parameters
\section{Parameters of the binary system}
\label{sec:sys}
Using the duration and the period of the eclipse, as well as the white dwarf mass constraints obtained in Sec.\,\ref{sec:constr}, we are able to derive precise orbital parameters for the corresponding binary system.

\subsection{Binary system equations}
\label{sec:equations}
The binary system is composed of a white dwarf of mass $M_{\rm wd}$ and a companion star of mass $M_{\star}$ and of radius $R_{\star}$. In IPs, there is evidence that the companion star is filling its Roche lobe, which means that its orbit should circularize on a short time scale \citep[e.g.][]{hurley2002}. Therefore, we assume that our system has a circular orbit with an orbital separation $a$ and an inclination $i$ compared to our line of sight (with $i=90$\degr\ corresponding to an edge-on system).

From Kepler's equations, we derive the following two relations,
\begin{equation}
	\frac{R_\star}{a} = \sqrt{\sin^2\left(\frac{\pi\, \Delta t}{T}\right) + \cos^2\left(i\right)},
	\label{eq:1}
\end{equation}

\begin{equation}
	M_\star +M_{\rm wd} = \frac{4 \pi^2 a^3}{G\,T^2},
	\label{eq:2}
\end{equation}

\noindent
where $\Delta t$ and $T$ refer to the duration and the period of the eclipse, respectively (Sec.\,\ref{sec:eclipse}) and $G$ is the gravitational constant. 

As already stated, the companion star is filling its Roche lobe, so we also use the approximation of \cite{eggleton1983} to describe its radius,

\begin{equation}
	\frac{R_\star}{a} = \frac{ 0.49 \, q^{2/3} }{ 0.6 \, q^{2/3} +\ln\left(1 + q^{1/3} \right) }, {\rm \hspace{0.5cm} where \hspace{0.5cm}}  q=\frac{M_\star}{M_{\rm wd}}.
	\label{eq:3}
\end{equation}

Finally, the donor mass and radius follow an empirical relation derived by \citet{patterson1984}. The corresponding relation is discontinuous at $M_\star\sim0.8$\msol\ but can be approximated by,

\begin{equation}
	M_\star \sim \frac{R_\star}{{\rm R}_\odot} {\rm M}_\odot.
	\label{eq:4}
\end{equation}

\noindent
This approximation is further justified by the comparison of the final parameters in Fig.\,\ref{fig:systemplot} and in Sec.\,\ref{sec:results}.

\subsection{\sr\ system parameters}
\label{sec:results}
\begin{figure}
	\centering
	\includegraphics[width=\columnwidth]{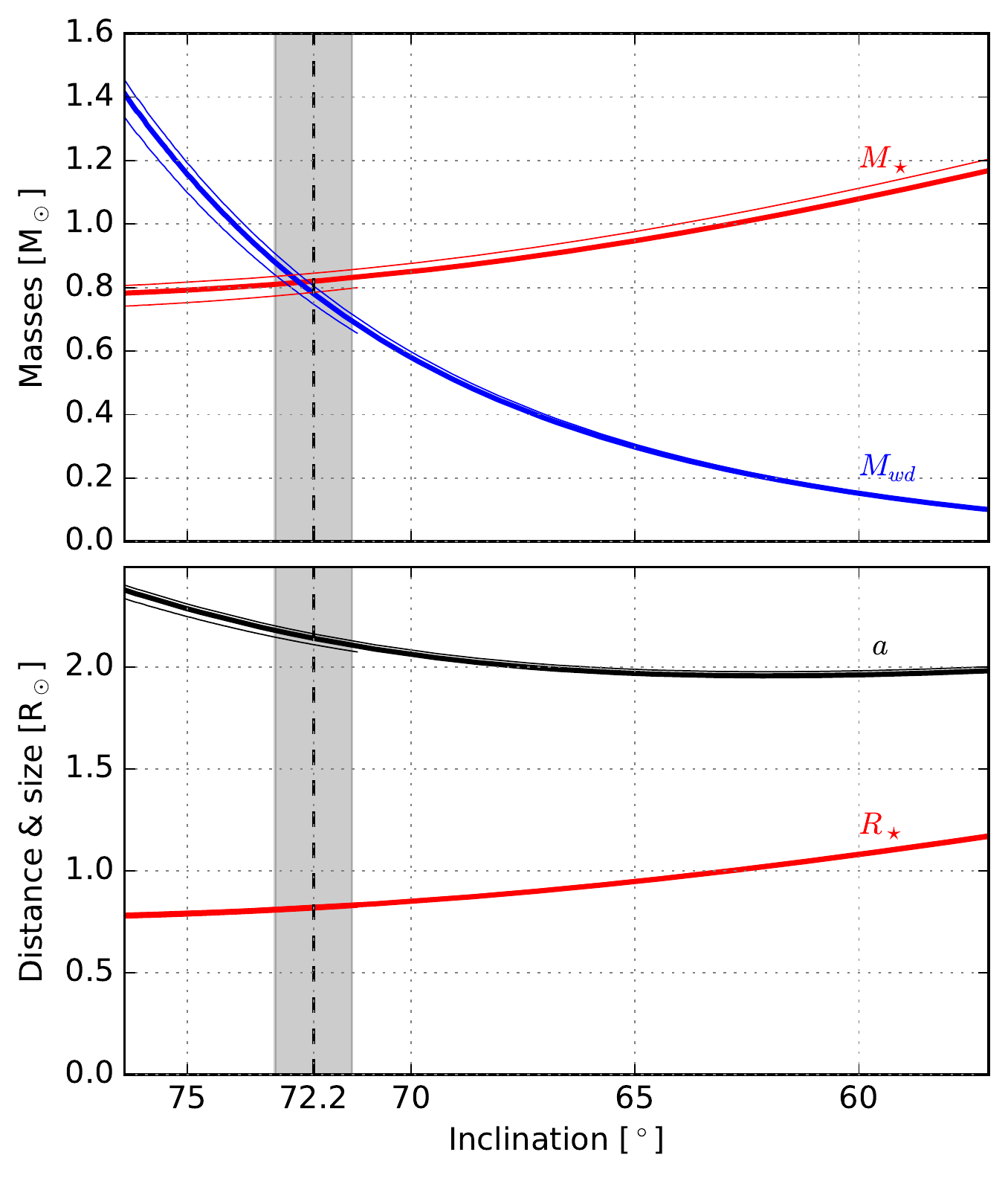}
	\caption{Parameters of \sr\ binary system derived from the duration and the period of the eclipse (Sec.\,\ref{sec:eclipse}), and presented as functions of the system putative inclination. Top panel: white dwarf mass (blue) and stellar mass (red). Bottom panel: stellar radius (red) and orbital radius (black). The parameter space was limited to the interval 0.1--1.4\msol\ for the white dwarf mass, and the measurement obtained from the spectral analysis is delimited by the grey shade (90\% confidence interval around the best fit value shown by the dashed line). For each parameter the thick line corresponds to the value derived using the mass-radius relation given in eq.\,(\ref{eq:4}) while the thin lines are obtained by using the discontinuous empirical relation given by \citet{patterson1984}. The results obtained are very similar and, in this figure, the thin lines are partly overlaid by the thick ones (e.g.\ for $R_\star$).  
	}
	\label{fig:systemplot}
\end{figure}

Using the parameters of the eclipse, $\Delta t$ and $T$, derived in Sec.\,\ref{sec:eclipse}, we numerically solved the system of equations introduced in Sec.\,\ref{sec:equations}. The results are presented as functions of the system inclination $i$ (Fig.\,\ref{fig:systemplot}). We restricted the parameter space to the standard white dwarf mass range, $M_{\rm wd} = 0.1$--$1.4$\msol, which corresponds to an orbital inclination $i=57.1$--$76.4$\degr, an orbital separation $a=2.0$--$2.4$\rsol, and a mass $M_\star = 0.75$--$1.2$\msol\ for the companion star.

The parameters of the \sr\ binary system can then be fully constrained by using the mass of the white dwarf obtained from our spectral analysis (see Table\,\ref{tab:param} and Fig.\,\ref{fig:systemplot}).  Thus, their precise values are: $M_{\rm wd}=0.78^{+0.10}_{-0.09}$\msol, $M_\star=0.82\pm0.01$\msol, $R_\star=0.82\pm 0.01$\rsol, $a=2.14\pm0.04$\rsol, and $i=72.2\pm0.9$\degr. The uncertainties listed here only account for the propagation of the errors on the white dwarf mass through the system of equations presented in Sec.\,\ref{sec:equations}.

The systematic errors derived for the orbital period and the eclipse duration do not significantly change the constraints we obtain on the system parameters, except for the orbital inclination. As the eclipse duration increases, the curves presented in Fig.\,\ref{fig:systemplot} shift towards higher inclination angles, leading to an inclination $i=74.6\pm1.0$\degr\ when the eclipse duration reaches its upper limit $\Delta t \leq 37.1$\,min.

We also tested the approximation we made by using eq.\,(\ref{eq:4}) instead of the empirical mass-radius relations given by \citet[][eq.\,3]{patterson1984}. We solved our system of equations (\ref{eq:1}, \ref{eq:2}, and \ref{eq:3}) using both parts of his relation: (i) $M_\star \leq 0.8$\msol\ and (ii) $M_\star >0.8$\msol, and the results are plotted in Fig.\,\ref{fig:systemplot}. As the relation of \citet{patterson1984} is discontinuous, both regimes (i) and (ii) can be verified for \sr. The parameters derived for the best fit white dwarf mass are within the error bars listed previously, except for $M_\star$. The mass of the donor would be 0.78 or 0.85\msol\ in regime (i) and (ii), respectively.

%%%%%%%%%% Section 4: Conclusion
\section{Discussion and Conclusions}
\label{sec:conclu}

Based on the orbital and spectral parameters we derived using the \nustar\ legacy program observations, we identified \sr\ as an eclipsing Intermediate Polar. The inclination we derived for this source, $i=(72.2^{+2.4}_{-0.0})\pm1.0$\degr, is consistent with the expectation that the system is  close to being edge-on, based on the eclipses we detected. In this case, any strong misalignment between the orbital inclination $i$ and the magnetic field inclination $\alpha$ would produce a self-occultation of the X-ray source at the white dwarf spin period\footnote{The X-ray emission is mainly created close to the magnetic poles of the white dwarf (see Sec.\,\ref{sec:model}).}. Therefore, the non-detection of spin modulation might be an indication that the orbit and the magnetic field inclinations are not far apart (i.e. the inclination of the magnetic field could be close to the upper limit derived from the spectral analysis, $\alpha<60$\degr).

The white dwarf mass $M_{\rm wd}=0.78^{+0.10}_{-0.09}$\msol\ of this system is close to the average values published for the brightest known IPs \citep[e.g.][]{suleimanov2005,yuasa2010,hailey2016}. The average X-ray luminosity of this bright population is $L_{0.1-100keV}\sim 2\times10^{34}$\ergs\ and \sr\ would reach this luminosity if located at a distance $d>3$\,kpc. The high Galactic absorption anticipated from the source spectrum is compatible with such a large distance and this is therefore strong evidence that \sr\ is not a Polar, since Polars tend to have lower accretion rates and therefore lower luminosities \citep[$ L_{3-20keV} < 10^{32}$\ergs,][]{sazonov2006}.
Furthermore, a distance of several kiloparsecs explains the relative faintness of \sr\ compared to the previously identified IPs for which $d<1$\,kpc \citep[][and references therein]{suleimanov2005}. In addition, the sample of known IPs is probably not representative of the whole population. Indeed, the properties of the faint population of magnetic CVs can be investigated through their putative contribution to the Galactic Ridge X-ray emission: an average mass of about 0.5\msol\ is anticipated for these unresolved sources in order to explain the level of emission measured in the hard X-rays \citep[][and references therein]{krivonos2007}. In this case, \sr\ would be on the higher end of the IPs' mass distribution, even if populations with masses larger than 0.9\msol\ have also been anticipated in the Galactic center region \citep{perez2015,hong2016} and if systems with masses close to the Chandrasekhar limit have been reported \citep[e.g.][]{tomsick2016}.

The donor star mass and radius, $M_\star=0.82\pm0.01$\msol\ and $R_\star=0.82\pm 0.01$\rsol, are compatible with the star being of type K, similar to those observed in other CVs having similar orbital periods \citep[e.g.][]{knigge2006}. The near-IR counterpart reported by \citet{karasev2012}\footnote{From the UKIDSS-DR6 Galactic plane survey catalog, the near-IR counterpart of \sr\ is observed in three bands (${\rm J}=16.750\pm0.016$, ${\rm H}=15.714\pm0.014$, ${\rm K}=14.356\pm0.011$) and is located at ${\rm R.A.}=18^{\rm h}29^{\rm m}20.16^{\rm s}$, ${\rm Dec.}=-12^\circ12'50.3''$ (J2000).} cannot be used to confirm this stellar type because, in IPs, this energy range is likely to be dominated by the accretion disk \citep{knigge2006}. However, it can provide an independent estimation of the column density towards this source. Assuming a flat intrinsic spectrum and using the conversion factor provided by \citet{cox2000}, we derived the column density $N_{\rm H}\sim2.6\times10^{22}$\,cm$^{-2}$. Such a high extinction is expected for a distant source close to the Galactic plane (Galactic latitude $b=-0.7$\degr), and is therefore consistent with what has been inferred from the X-ray observations.

By improving the completeness of the hard X-ray Galactic faint sources, the \nustar\ legacy program `Unidentified \integral\ sources' also aims at improving our knowledge of the luminosity functions of the different categories of sources and in particular to address whether there is a faint HMXB population. In this context, \sr\ was successfully excluded from this putative population, and several similar observations are being made in order to help the identification of additional faint persistent sources. They will be the subject of future publications.

\section*{Acknowledgments}
This work was supported under NASA Contract No. NNG08FD60C, and made use of data from the \nustar\ mission, a project led by  the California Institute of Technology, managed by the Jet Propulsion  Laboratory, and funded by the National Aeronautics and Space Administration. We thank the \nustar\ Operations, Software and  Calibration teams for support with the execution and analysis of these observations.  This research has made use of the \nustar\ Data Analysis Software (NuSTARDAS) jointly developed by the ASI Science Data Center (ASDC, Italy) and the California Institute of  Technology (USA). RK acknowledges support from Russian Science Foundation (grant 14-22-00271). GP acknowledges the Bundesministerium f\"ur Wirtschaft und Technologie/Deutsches Zentrum f\"ur Luftund Raumfahrt (BMWI/DLR, FKZ 50 OR 1408).

\bibliographystyle{mn2e}
\bibliography{Clavel_IGRJ18293-1213_accepted}

\label{lastpage}

\end{document}